# Bond dissociation dynamics of single molecules on Ag(111)

Donato Civita[1], Jutta Schwarz[2], Stefan Hecht[2], and Leonhard Grill[1*]

[1]Department of Physical Chemistry, University of Graz, Austria
[2]Department of Chemistry and IRIS Adlershof, Humboldt-Universität zu Berlin, Germany

**Abstract**

The breaking of a chemical bond is fundamental in most chemical reactions. To understand chemical processes in heterogeneous catalysis or on-surface polymerization the study of bond dissociation in molecules adsorbed on crystalline surfaces is advantageous. Single molecule studies of bond breaking can give details of the dissociation dynamics, which are challenging to obtain in mole-scale ensemble experiments. Bond breaking in single adsorbed molecules can be triggered using the energy of the tunnelling electrons in a scanning tunnelling microscope (STM)[1] at selected positions[2] to investigate the dissociation dynamics.[3] Single bond dissociation dynamics has been deeply investigated only in small molecules,[4] but not in larger molecules that exhibit distinct rotational degrees of freedom. Here, we use low temperature (7 K) STM to dissociate a single bromine atom from an elongated molecule (dibromo-terfluorene) adsorbed on a Ag(111) surface. This rod-like molecule allows to clearly identify not only displacement of the reaction fragments, but also their rotation. The results show that the molecular fragment binds to the nearest silver atom and only further rotation is allowed. Moreover, the excitation responsible for the bond breaking can propagate through the molecular backbone to dissociate a bromine atom that is not located at the pulse position. These results show the important role of the metal substrate in conditioning the bond dissociation dynamics. Our results might allow to improve the control of the synthesis of 2D materials and targeted engineering of molecular architectures.

In a chemical reaction, the characterization of bond breaking dynamics aims to understand how the initially supplied activation energy is redistributed from the parent to the product. molecules. Unveiling the underlaying chemical energy conversion mechanism allows a better understanding and therefore control of chemical reactions. To characterize the dynamics, various parameters of a chemical reaction must be known: the energy and collision geometry of the reactants as well as the energy and angular distribution of the products. Many insights have been obtained with crossed molecular beam experiments[5,6] where two well-defined beams of atoms or molecules collide in vacuum. However, this method integrates over many possible geometries, resulting in a distribution of directions and energies where the dynamics of a single reaction cannot be assigned. This is different in so-called surface-aligned reactions[7,8] where the reactants are positioned on a metallic single-crystal surface. Metal surfaces are fundamental in heterogeneous catalysis as they lower the activation energy of the reaction.[9] This function is not only used in heterogeneous catalysis where the products typically leave the surface after the reaction, but also for the synthesis of covalently bonded networks.[10,11] In these reactions, halogenated precursor molecules (monomers) adsorb on a metal surface, which subsequently catalyses halogen dissociation to obtain an activated precursor. Therefore, the study of molecular dehalogenation on crystalline metal surfaces is important to understand and control such polymerization process and the synthesis of two-dimensional architectures.

For detailed insight into bond dissociation dynamics on a surface, it is advantageous to investigate the reaction at the single molecule level. Single bond dissociations have been triggered with electron pulses from the tip of scanning tunnelling microscope (STM).[1,2,12–16] In particular, the STM technique has been used to study the carbon−halogen bond dissociation of halogenated hydrocarbons.[17–20] Specific bonds within an individual molecule can be cleaved without influencing other bonds by choosing an appropriate bias voltage and position of the electron pulse.[1,2,12] Typically, only one bond is broken, but in rare cases a single electron event can dissociate two bonds within a molecule, or a chain of molecules, due to charge delocalization.[21–23] Despite all these studies, there is a lack of experiments that investigate the rotation of fragments as well as the effect of different pulse positions for the same bond dissociation. Small molecules are disadvantageous in this regard since different pulse positions affect the same area of the initial compound and the orientation of the fragments is difficult to recognize in STM images. Therefore, it is challenging, if not impossible, to identify the orientation and subsequent rotation of small fragments. Nevertheless, this property is important to describe how the energy, released during the bond dissociation process, is being redistributed throughout the molecule and the cleaved fragment (halide ion) in magnitude and space, in order to understand directional properties of chemical bonds.

In order to have access to the rotation of molecular fragments, we used elongated dibromoterfluorene (DBTF) molecules for the study of single-bond dissociation dynamics. They are composed of three fluorene groups, which constitute the elongated backbone of the molecule (Fig. 1A). Each fluorene group has lateral dimethyl groups, which lift and decouple the molecule from the substrate. At the two extremities, bromine substituents are attached.[24] In our experiments, we selectively induce dissociation of one of the terminal Br substituents from the terfluorene backbone by using a pulse of tunnelling electrons from the STM tip. Thanks to its rod-like shape, the orientation of the molecule can be unambiguously identified. In this way, the molecule can be positioned on the surface with a defined orientation and can be imaged after debromination in its final position and orientation. Moreover, distinct tip positions for the electron pulse can be explored.

DBTF molecules were deposited under ultrahigh vacuum (UHV) onto the clean (111) surface of a silver single-crystal at room temperature, and studied by scanning tunnelling microscopy after transferring and cooling the sample in the same UHV chamber to 7 K. The molecules were found intact and self-assembled in islands. A single molecule was then extracted from the island by lateral manipulation with the STM tip. The STM image of a single DBTF molecule, presented in Fig. 1B, shows three dominant lobes with 1.60 ± 0.03 Å apparent height, which originate from the dimethyl side groups. Additionally, the Br substituents appear as shoulders at the two extremities, which are about 2.8 nm apart.

The orientation (angle) of isolated intact DBTF molecules relative to the substrate was measured for 199 different molecules. The resulting data, shown in Fig. 1C, gives an orientation of 4.5° ± 1° of single adsorbed DBTF molecules with respect to the $[1\bar{2}1]$ high symmetry direction of the Ag(111) surface. Moreover, with the help of bromine atoms adsorbed nearby a single DBTF molecule, we extracted the precise adsorption position of the molecule with respect to the atomic lattice of the substrate. This analysis leads to the DBTF adsorption configuration on Ag(111) surface shown in Fig. 1D.

We proceeded then with the dissociation of a C-Br bond within a single DBTF molecule by electron pulses from the STM tip, as shown in Fig. 2. After imaging an intact DBTF molecule, the STM tip was placed at a specific position above the molecule (indicated by circles in Fig. 2A), and the tip-sample distance was set constant. Then, a voltage bias of 2.0 V was applied to the sample (the tip was grounded). At the same time, the bond dissociation was identified as a single discontinuity in the current-vs-time trace that was recorded during the experiment. As soon as this single jump of current had been observed, the bias was set to zero to prevent further reactions or displacements of the molecule. Indeed, voltage pulses that show multiple jumps in the current trace were discarded.

The products of the bond dissociation, *i.e.* the isolated Br ion and the molecular fragment bromoterfluorene (BTF), were observed in the subsequent STM images (Fig. 2B and C). The Br atom (dashed circles) appears in a characteristic hat-shape, known for halogen atoms adsorbed on metal surfaces.[25] The bright feature with three lobes is the singly de-brominated BTF molecule. Due to the chemical change caused by the bond dissociation, the de-brominated molecule has a slightly different appearance. The change in shape is clearly visible when comparing height profiles of the intact and de-brominated molecule (Fig. 2D) as the Br shoulder of the intact molecule (solid line) is missing after the electron pulse. Instead, a depression appears on the de-brominated side (dashed line), which is probably an electronic feature due to the molecule-substrate interaction.[13,26,27] Moreover, the dimethyl lobe next to the de-brominated side (see arrow in Fig. 2D) has a reduced height after dissociation. This observation indicates a stronger interaction of the de-halogenated side of the molecule with the silver surface. Indeed, after bromine dissociation, the BTF fragment is probably left with an unpaired electron on the carbon atom constituting a surface-stabilized radical,[19,28–33] which is pinned via a single binding point to the metal surface. The enhanced interaction of this end of the BTF fragment with the surface is verified by lateral manipulation experiments, from which the stronger molecule-surface binding is evident (Fig. 5A).

For deeper insight into the dissociation dynamics, we repeated the dissociation experiment a number of times, each time with a new intact molecule. Experiments where voltage pulses did not lead to dissociation or where multiple jumps are observed in the current trace were excluded. The set of dissociations are grouped into three distinct cases, which differ from the pulse position above the molecule. The three pulse positions are the left bromine shoulder, the central lobe, and the right bromine shoulder, indicated in Fig. 2A by red, black and blue circles, respectively. 32 successful dissociations were achieved by pulsing on the left Br shoulder, 59 on the central lobe, and 26 on the right Br shoulder. Due of the symmetry of the molecule, the pulses on the left and right Br shoulder might appear equivalent. However, they are not, because the molecule is aligned asymmetrically with respect to the substrate atomic lattice. Specifically, there is a mismatch angle between the molecular long axis and the $[1\bar{2}1]$ high-symmetry direction of the Ag(111) surface, and the length of the molecule is not commensurate with the lattice. In Fig. 1D, it can be clearly seen that the two C-Br bonds on either end have different local surroundings of the underlying atomic lattice.

Voltage pulses typically affect the molecule locally, due to the spatial confinement of the tunnelling electrons.[1,2,12] Accordingly, one might expect that only intramolecular bonds that are located directly underneath the STM tip can be dissociated. However, we observe that, for any of the three pulse positions, either the left or the right C-Br bond can dissociate, even the bond that is most distant from the pulse position. The probabilities of left and right bond dissociation are indicated with arrows in the centre of Fig. 2, with different colours depending on the pulse position. When the tip is placed above one of the two Br shoulders, the probability to break the C-Br bond underneath is highest (79% or 71% if pulsing on the left or right shoulder, respectively). Nevertheless, the dissociation of the C-Br bond at the opposite side of the molecule has still remarkably high probabilities of 21% or 29%, despite the relatively large distance of 2.9 nm. Additionally, pulses at the centre of the molecule can cause both left or right Br dissociation, with equal 50% probability.

We investigated the mechanism underlying the dissociation at the rather remote distance of 2.9 nm from the pulse position. To understand whether the process is mediated by the molecule or not, we applied equivalent electron pulses at the same distance from the C-Br bond (2.9 nm), but on the bare silver surface around the molecule. Despite the equal distance from the STM tip, in no case did these pulses trigger any Br dissociation. Electric field and surface-mediated excitations, which could in other cases explain remote tip-induced bond dissociations,[13,34,35] can thus be excluded here. Clearly, the excitation propagates *through* the molecule, from the position where the tunnelling electrons are injected to the bond that breaks. One way to induce single bond dissociation with an STM tip is the excitation of molecular vibrations via inelastic electron tunnelling.[1] However, this excitation pathway does not appear to be the case here since the process is not symmetric for positive and negative bias polarity.[15] Instead, C-Br dissociation can only be induced by positive voltage pulses (applied to the sample, while the tip is grounded). Additionally, the threshold voltage required for Br dissociation, 1.85 V, is much higher than typical vibrational energies (60-400 meV).[36–38] We therefore relate the C-Br bond dissociation to resonant electron tunnelling into an unoccupied electronic state, as observed before.[29,39] This electronic state might be rather delocalized over the entire molecule and this might explain why bond dissociation can occur rather remote to where the initial electron pulse is applied.[21,24]

In order to get insight into the molecular dynamics, it is important that – during a voltage pulse – the molecule does not rotate or move before the dissociation. This is guaranteed since we discard experiments with multiple jumps in the current trace. In the experiment, the position and orientation of the intact molecule with regard to the substrate lattice were first identified from the STM image before dissociation. Then, the position of the two products, *i.e.* bromine atom and BTF molecular fragment, and orientation of the BTF fragment, were determined from the image of the same surface area after dissociation. The position of DBTF or BTF is defined via the maximum of the central lobe, their orientation as the angle between the horizontal direction and the line passing through the two side lobes. Positive angles were measured counter-clockwise. These data were collected for each dissociation experiment. Position and orientation of the intact molecule served to align all dissociation experiments in the same reference frame. From the position data we were able to obtain the spatial distribution of the products, reported in Fig. 3A and D for the left and right bond dissociations, respectively (the starting condition is represented by the molecular structure drawn in grey in the background). In addition, the final orientations of the BTF fragment are plotted in Fig. 3B-C. Analysing the scatter plots of Fig. 3A and D for the three pulse positions (see the different colours) we observe that the datapoints follow a similar distribution regardless of the pulse position. Therefore, as a first important conclusion, the dissociation dynamics are independent on the electron injection position on the molecule.

For the left bond dissociation (Fig. 3A) the Br atoms are distributed exclusively on the left side of the plot, while for the right dissociation they are only on the right side (Fig. 3D). This shows that, upon dissociation, the Br moves away from its initial position (see the molecular structure in Fig. 3A and D), roughly following the prior C-Br bond direction, similar to what has been observed before.[40] However, the Br position distribution has a broad spread in angles (about ± 60° for the left dissociation and ± 30° for the right dissociation) and recoil distance (± 3 Å with a mean value around 5 Å in both left and right dissociation). This wide distribution of the Br atoms might be caused by rotations of the intact molecule upon bond dissociation. In such a case one would expect a correlation between Br positions and orientations of the BTF fragment after dissociation. However, such a correlation is not observed as visible via the colour-coding in Fig. 3A (see Fig. 4 upper panel). This result suggests that the Br atoms are rather deflected by interaction with silver atoms of the surface after the dissociation.[4]

In contrast to the Br atoms, the distribution of the BTF position groups into defined clusters of data points. We observe six clusters for the left bond dissociation (Fig. 3A), and two clusters for the right dissociation (Fig. 3D). Considering the initial position of the DBTF molecule before bond dissociation (*i.e.* the origin of the coordinate systems in Fig. 3A and D), the BTF fragment has moved by about 1 Å and 2 Å for the right bond dissociation, while for the left bond dissociation larger distances between about 1 Å and 8 Å are observed. However, these larger distances from the origin do not correspond to translations but rather rotations of the molecular fragment. In fact, for the left bond dissociation, by inspecting the histogram of the BTF orientation angles (Fig. 3B) it becomes clear that in 66% of the cases the fragment undergoes substantial rotational movement. These rotations are clockwise and reorient the BTF fragment to -2.5° and -27.0°. This preferential (clockwise) verse of rotation is similar to what has been observed for diiodobenzene on Cu(110).[20] In the other 34% of the left dissociations the BTF orientation remains essentially unchanged (about 1° rotation). In strong contrast, no relevant rotations of the fragment were observed for the right dissociations (see the only peak in Fig. 3C, very close to the original orientation).

The case of left bond dissociation allows us to investigate the correlation between the three orientations with the positions of the BTF fragments (Fig. 4). Interestingly, we observe that couples of clusters of BTF positions (triangles in Fig. 4A) are related to specific final orientations (Fig. 4B). Therefore, it appears that the rotation of the fragment explains the displacement of its centre from the origin.

Regarding the difference observed between left and right dissociations, the role of the pulse position can be excluded after the analysis of the colour-coded scatter plots of Fig. 3A and D. Moreover, the left and right Br substituents are chemically equivalent and cannot play any role. Therefore, we assign the observed difference between left and right dissociations to the adsorption configuration. Specifically, the different alignment of the two C-Br bonds with the respect to the Ag(111) lattice creates a unequal chemical surrounding. In this way, the nonsymmetric adsorption seems to determine the rotations or translations observed in both cases.

To gain even deeper insight into the dissociation dynamics, we investigated the state of the de-brominated BTF fragment and found that the BTF can only rotate around a fixed pivot point (Fig. 5A). The pivot point can be determined for each dissociation experiment and is plotted in Fig. 5B and C for the left and right dissociations, respectively. For both the left and right bond dissociations, the observed pivot points split into only two clusters, localized around the position of surface silver atoms. These silver atoms are the nearest to the C-Br bond that dissociates (as seen from overlapping molecular structure and surface lattice in Fig. 5), showing that after dissociation the de-brominated side of the molecule interacts with the nearest silver atoms, allowing only rotation of the fragment. This strong interaction is probably caused by the free electrons of the metal surface that readily couple to the unpaired electrons of the de-brominated side of the molecule.[26,27,32] As result, the BTF fragment becomes a surface-stabilized radical with a binding position at the surface defined by the silver lattice.

Considering momentum conservation, one might expect that the BTF fragment moves always in opposite direction to the Br atom. Interestingly, this is not the case for both left and right dissociation. Indeed, the two binding positions indicate a "backward" and also a "forward" recoil of the BTF fragment, toward the Br atom (see the cluster of crosses indicated with "bwd" and "fwd" in Fig. 5B and C). The "forward" recoil cases are not rare, but amount to 30% and 79% for the left and right bond dissociation, respectively. Hence, the BTF can move, counterintuitively, in the same direction as the Br atom, to reach one of the two binding positions. This observation shows that the molecule-surface interaction can strongly affect the dissociation dynamics on metal surfaces.

Finally, we analysed the positions of the dissociated Br atom in relation to the binding point of the BTF fragments by colour-coding the datapoints and also the orientation histograms. The resulting plots are shown in Fig. 4C and D for the left dissociation (similar plot for the right dissociation). We observe that the position of Br atoms (circles) is not related to the binding point of BTF fragments (crosses). Moreover, from Fig. 4A and B the position of the Br atom has also no relation with the rotation of the molecular fragment. Therefore, we conclude that right after the bond dissociation, the two products follow uncorrelated dynamics.

In conclusion, we studied electron-induced C-Br bond dissociation in DBTF molecules on Ag(111). We found that the excitation responsible for the bond breaking can also be mediated by the molecular backbone to cleave the C-Br bond at almost 3 nm distance from the pulse position. Our experiments clearly show that the surface can steer the dissociation dynamics. Right after C-Br bond cleavage, the dynamics of Br atom and BTF fragment are uncorrelated. It appears that the Br atom is ejected along the prior bond direction, and the interaction with the surface atoms deflects its trajectory. On the other side, the BTF fragment binds its de-brominated side to one of the nearest silver surface atoms, which acts as pivot point. The kinetic energy still available for the fragment can eventually be dissipated in rotation. Furthermore, the dynamics of the dissociation do not depend on the position of the current injection into the molecule. We believe that our results pave the way towards improved understanding of surface-catalysed reactions.

**Acknowledgements**

We gratefully acknowledge the financial support from the "Doc Academy NanoGraz" at the University of Graz, the European Commission via the MEMO project (FET open project no. 766864), and the Austrian Science Fund FWF via the CCC project (international project no. I 4897).

**Figures**

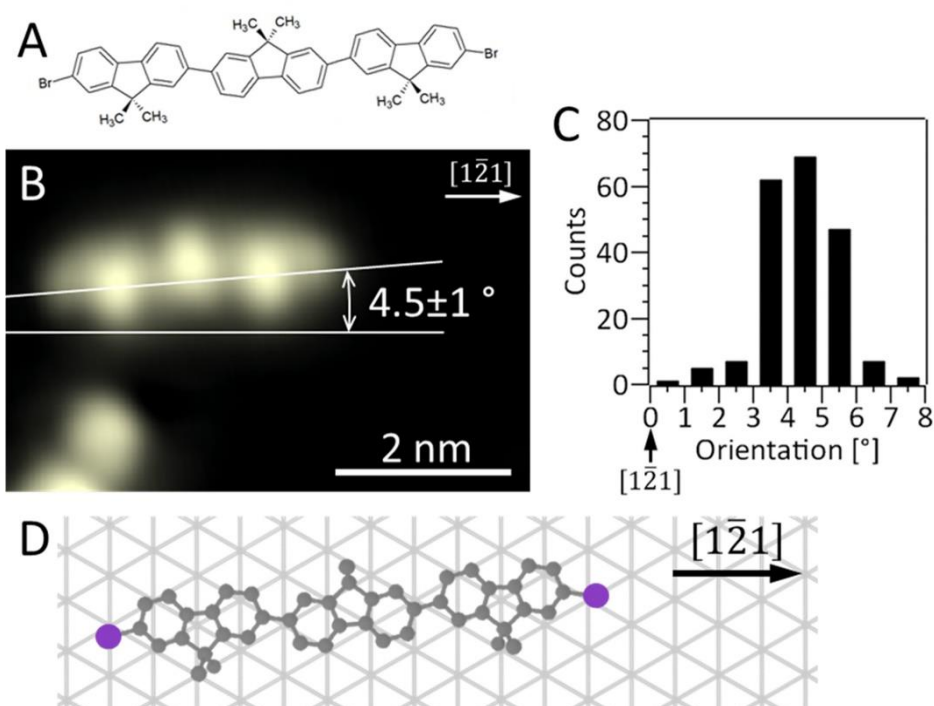

**Figure 1. Adsorption configuration of a DBTF molecule on Ag(111).** (A) Chemical structure of dibromo-terfluorene (DBTF). (B) STM image (0.6 V, 300 pA) of a single DBTF molecule on Ag(111). The orientation angle of the molecule with respect to the [1$\bar{2}$1] direction of the surface is indicated. (C) Histogram obtained from 199 images of single DBTF molecules. The data are folded considering the 6-fold symmetry of the Ag(111) surface and the symmetry of the molecule. In this histogram the [1$\bar{2}$1] symmetry direction of the Ag(111) surface is set at 0°. (D) Scheme of the adsorption configuration of a single DBTF molecule on Ag(111), obtained from an STM image of the molecule surrounded by six Br atoms. Grey circles represent carbon atoms and the violet circles bromine atoms, while the silver surface is plotted as a lattice.

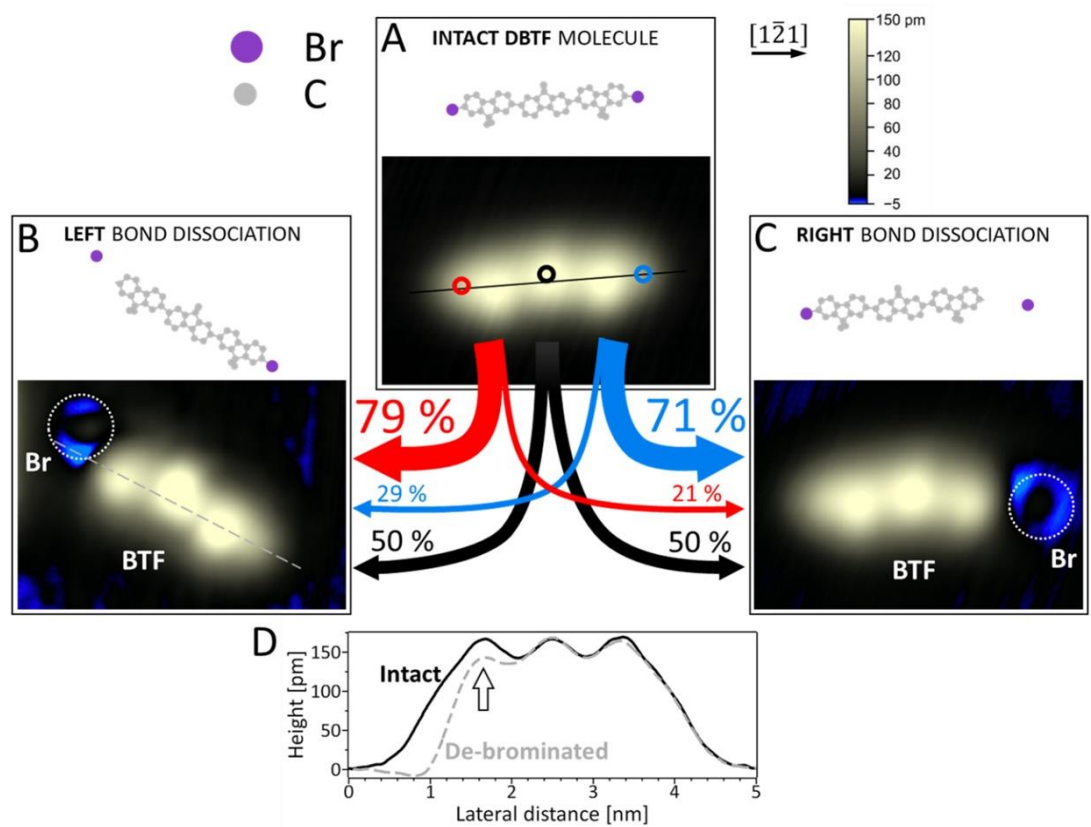

**Figure 2. Electron-induced dissociation of a Br atom from a DBTF molecule on Ag(111).** STM images (each 5.0 × 3.5 nm² in size, 300 pA, 0.6 V) of single DBTF molecules on Ag(111) before (A) and after (B,C) Br dissociation with corresponding schemes above them (grey and violet circles represent C and Br atoms, respectively). The colour scale is shown on the top-right corner, blue colour indicates a depression below the black level of the substrate set at 0 pm. The blue "hat-shaped" feature in the dashed circle (B,C) is the Br atom. All images are oriented equivalently, the $[1\bar{2}1]$ high-symmetry direction of the surface is indicated on the upper right. The red, black and blue circles in (A) indicate the positions of voltage pulses from the STM tip. The two STM images in B and C are obtained from two distinct dissociation experiments, and show the two possible outcomes. The coloured arrows show the measured probability of left and right dissociations at each pulse position. 117 molecules were dissociated in total: 32 with pulse on the left Br shoulder (red), 59 on the central lobe (black), and 26 on the right Br shoulder (blue). (D) Height profiles along the intact and the de-brominated molecule, their positions are indicated in (A) and (B), respectively.

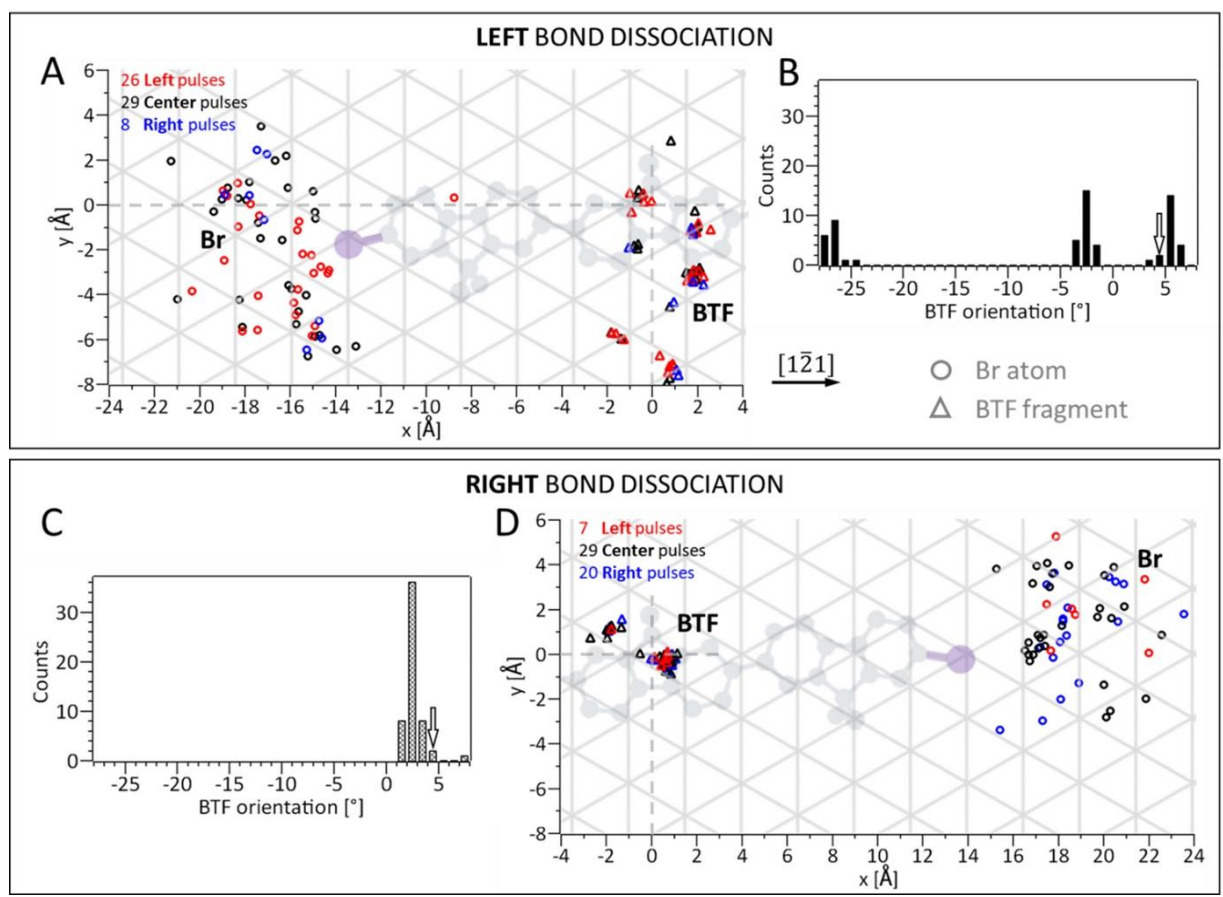

**Figure 3. Position and angle distribution of the fragments.** (A,D) Position distribution of the Br atom and BTF molecular fragment and (B,C) histograms of the BTF orientation, for the left and right bond dissociation (upper and lower panel, respectively). Due to the 6-fold Ag(111) surface symmetry, the datapoints are folded. In A and D, for each dissociation experiment, the central lobe of the intact molecule is set at the (0,0) position of the plot, and the molecule is oriented at +4.5° with respect to the [1$\bar{2}$1] direction of the Ag(111), which is horizontal. The chemical structure of the intact molecule is reported in grey (with the Br atoms in purple), and the Ag(111) surface is plotted as a grey lattice, where each crossing point represents a silver atom. Positions of the Br atom (circles) and positions of the central lobe of the BTF fragment (triangles) are indicated. The three different colours indicate the position of the electron pulse (as in Fig. 2A). In B and C there are the histograms of the orientation angles of the BTF fragment, where the [1$\bar{2}$1] direction is at 0°. The arrows show the orientation of the intact molecule.

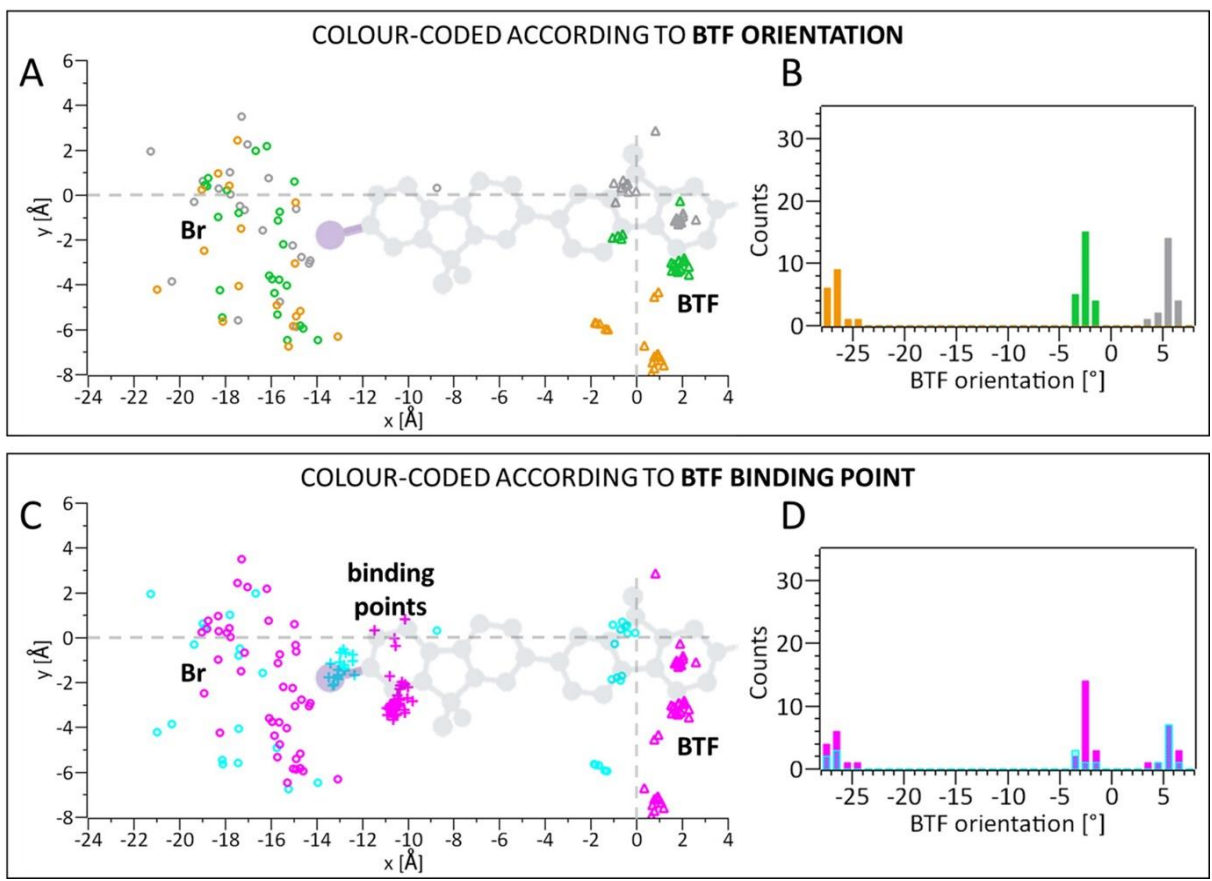

**Figure 4. Colour coded data.** In the upper and lower panel there are the same scattering plot and histogram of Fig. 3A and B. In the scattering plot the circles and the triangles represent the position of the bromine and molecular fragment, respectively. In C also the binding points are indicated with crosses, as obtained from Fig. 5. The intact molecular structure is also shown in its original configuration. The histogram shows the orientation after dissociation of the molecular fragment. Here the data points and histogram bars are colour-coded. In the upper panel, the colours are relative to the three different peaks of the histogram in B, which indicate the three possible final orientation of the BTF fragment (a similar representation of the right bond dissociations is trivial, since there is only one orientation for the BTF). In the lower panel, the colours are relative to the two different cluster of BTF binding points of the scatter plot in C (similar representation can be made for the right dissociations).

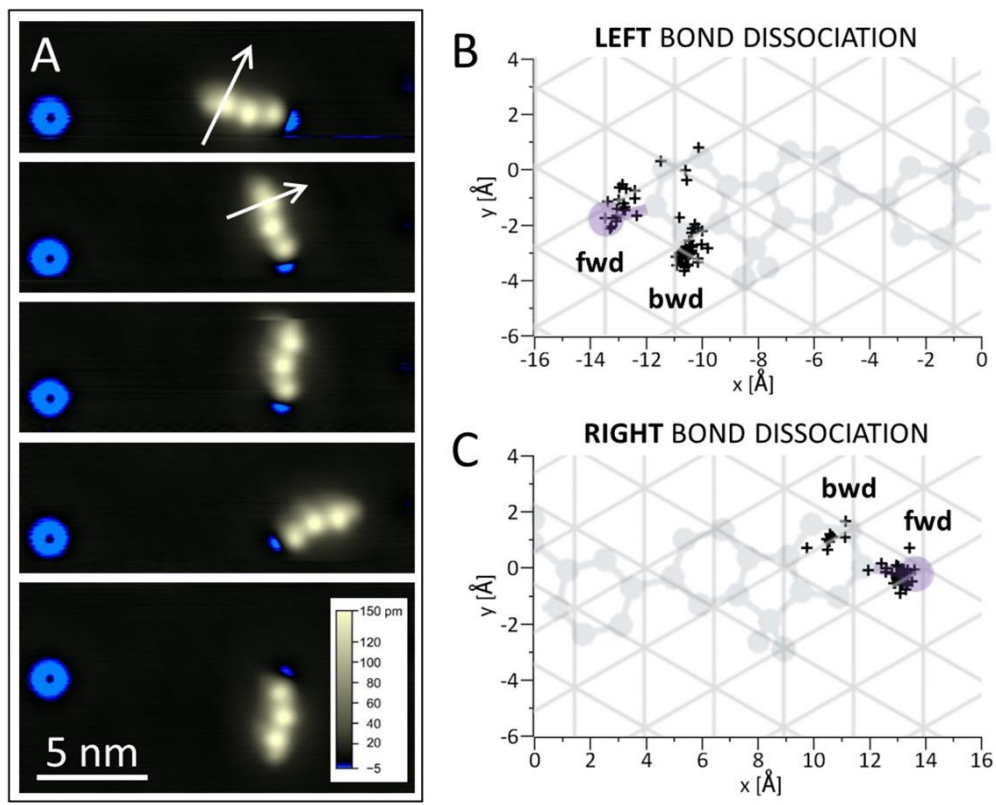

**Figure 5.** The panel in A sows a series of STM images (0.2 V ,500 pA), each recorded after a tip lateral manipulation (0.1 V, 80 nA) of a de-brominated DBTF molecule. The molecular fragment is on the right and is recognizable from the three bright lobes with a blue depression on one side. The Br atom is on the left and appears as a hat-shaped blue depression that remains at the same position as reference. On the first two images the white arrow indicates the path of the tip during the lateral manipulation. After each manipulation, the de-brominated DBTF molecule rotates around a pinned point. This set of images is used to extrapolate the position of the pinned point by overlapping them. B and C show the scatter plot of the pinned point position (crosses) extrapolated for each dissociation experiment, for the left and right dissociation, respectively. For both left and right dissociation, there are only two clusters of binding positions, indicated with "fwd" (forward) and "bwd" (backward). The molecular structure of the intact molecule and the substrate lattice are reported as reference to the initial conditions.